\documentclass[twocolumn]{aastex6}

\newdimen\figurewidth
\newcommand{\iiint}{\int}

\usepackage{amssymb}

\begin{document}

\title{A kinematic spiral arm shock signature: \\ ``Ringing" in the vertical motion of stars}

\author{Nir J. Shaviv\altaffilmark{1,2}}
\affil{$^{1}$Racah Institute of Physics, Hebrew University of Jerusalem, Jerusalem 91904, Israel \\
$^{2}$School of Natural Sciences, Institute for Advanced Study, Princeton, New Jersey 08540, USA}

 
\begin{abstract}

We analyze the motion of stars in the direction perpendicular to the galactic plane following a spiral arm passage. We show that the fast change in the vertical galactic potential causes a thermalized distribution to develop a distinctive  ``ringing"-like non-thermal signature.  We use A and F-stars from the extended Hipparocos catalogue to show that a spiral arm passage (or passages), with an amplitude (or randomly combined amplitudes) of at least $\delta \rho / \rho \gtrsim 0.15$ must have taken place in the past (with $\rho$ being the total background density). Presently, the local stellar distribution within $\sim 100$~pc of the plane, appears (at the 2.5$\sigma$ level) to be contracting towards it.   

\end{abstract}




%
%
%
%
%
%
%
%
%
%
%
%

\section{Introduction}

Our vantage point close to the galactic plane prevents us from directly seeing the spiral arms of the Milky Way. Nevertheless, the spiral structure can still be studied through the observations of various arm tracers  \cite[e.g.][]{Georgelin:1976,Blitz:1983,Vallee:1995,Dame:2001}. The arms can also leave kinematic signatures as they are well described by the \cite{Lin:1964} density wave theory \citep[e.g.,][]{Binney:1987,Lin:1995}.  For example, the local spiral arm dynamics can be derived from the position of the Hyades, Sirius and other moving groups \citep{Quillen:2005,Antoja:2011,Pompia:2011}. A similar signature can be derived from the Radial Velocity Experiment (RAVE) survey \citep{Siebert:2012}. Since the density waves can have a vertical component to the oscillation \citep{Debattista:2014,Faure:2014}, spiral arms could lead to a detectable vertical motion of stars, or masers \citep{Bobylev:2015}.

Another interesting aspect of the density waves is the behavior of the gaseous component. Although the overall density variations associated with the spiral arms are small, probably of order 10\%-15\% \citep[e.g.][]{Siebert:2012,Monguio:2015}, this is not the case with the gas. Since the speed of sound in the latter is typically significantly smaller than the pattern speed of the spiral waves, shock waves are easily formed. As a consequence, one can expect large abrupt jumps in the density \citep{Roberts:1972,Binney:1987}, which are in fact the perturbations that trigger star formation. Here we show that this abrupt change in density gives rise to an additional kinematic signature associated with the spiral arm passage, that of vertical ``ringing" of the stellar distribution.

We will also see below that finding kinematic signatures is interesting, not only because it could teach us about the nature of the spiral arms, but also because it has interesting ramifications to the understanding the average mass density in the galactic disk. The standard method to recover this density is to analyze the distribution of stellar coordinates and velocities in the direction perpendicular to the galactic plane  \citep{Oort:1960}. If one assume that a given tracer population of stars is kinematically relaxed, or at least phase mixed, then relations between the distributions can be used to derive the potential and total mass density, with typical values ranging from 0.07 to 0.26 M$_\odot$/pc$^3$ ~\citep{Bahcall:1985,Stothers:1998,Holmberg:2000,korchagin:2003}. 

When the estimated value for the total mass is compared with  estimates for the total baryonic matter (i.e., stars, remnants and ISM, \citealt{Holmberg:2000,Garbari:2011,McKee:2015}), the difference can be used to infer the amount of dark matter material.  Given that the \cite{Holmberg:2000,korchagin:2003} studies for the total amount of mass is based on the Hipparcos data set, they are generally considered the most reliable. The value found by \cite{Holmberg:2000} for the total amount of matter is 0.102$\pm$0.010 M$_\odot$/pc$^3$, and 0.095 M$_\odot$/pc$^3$ for the visible matter, leaving very little for unaccounted dark matter.  Similarly, \cite{korchagin:2003} find a total column density of $10.5\pm 0.5$\,M$_\odot$/pc$^2$ within 50 pc from the plane, giving a similar density. 

Nevertheless, having a larger amount of dark matter at the galactic plane is interesting as it may have interesting repercussions. For example, a higher density concentrated near the plane could imply that dark matter can self-interact and cool (leading to a ``double disk dark matter"), with interesting implications to our understanding of high energy physics \citep{Fan:2013}. A larger density at the galactic plane would also immediately explain the prominent 32 million year oscillations observed over the past 550 million years \citep{Shaviv:2014}, and possibly apparent periodicities in cometary impacts (e.g., \citealt{Randall:2014} and references therein). Thus, it is very interesting to consider effects that could distort the estimates for the total amount of mass in the plane. 

In their study, \cite{Pestana:2010} claimed that the vertical distribution of stars and their velocities is not in a state of equilibrium such that kinematic approaches to estimate the total amount of mass are futile. This point was later addressed by \cite{Salcedo:2011}, who argued that there are no inconsistencies with the assumption of a virial equilibrium, which in fact, probably holds. However, even if the distribution of stars is in virial equilibrium, we will show that the passage of a spiral arm can perturb the vertical distribution of stars, giving rise to a ``ringing" effect, such that it has the appearance of being in an equilibrium, but the vertical gravitational potential inferred from it would be wrong. We do so in \S\ref{sec:kinematicsignature}. The signature of this ringing is then parameterized in \S\ref{sec:asymetry}. In \S\ref{sec:nonImpulsive} we study the behavior of stars with a finite amplitude in response to the spiral arm passage, for which the perturbation is neither instantaneous nor adiabatic. We then show in \S\ref{sec:realdata} that a small ringing signature is present at the data, indicating that the stellar population in the solar vicinity is presently contracting towards the mid-plane. We end with a discussion in \S\ref{sec:discussion}.


\section{The kinematic signature of a spiral arm passage}
\label{sec:kinematicsignature}

Let us consider a simple model to describe the effect of a spiral arm passage. First, we approximate the vertical potential to be harmonic. This is a good approximation as long as the density, $\rho(z)$, can be assumed to be constant, namely, that it falls slowly when compared to the amplitude of the vertical oscillations of the stellar population. 

Furthermore, we assume that the spiral arm can be considered as a contribution of two components. The first  is the harmonically varying density of the background stars. Because its period is at least 100 Myr if not longer \citep{Shaviv:2002}, it can be approximated as an adiabatic variation of the vertical oscillation, which has a half period of 30 to 42 Myr \citep{Bahcall:1985,Stothers:1998,Holmberg:2000}. The second component is that of the ISM gas which has a shock wave and a slow rarefaction behind it. 

The effect of the adiabatic variation is to slowly puff up and contract the stars. For the harmonic oscillator, the adiabatic invariant is the area $2 \pi E/\omega$ enclosed in the $z-p_z$ phase space of a star. Since $\omega \propto \rho^{1/2}$, $E \propto \rho^{1/2}$, the amplitude of the oscillations will vary as $z_{max} \propto \rho^{-1/4}$. However, even though the parameters describing the stellar motion change, a thermalized distribution of stars will keep its thermalized form, albeit in a changed harmonic potential. On the other hand, the effect of an instantaneous increase in the density associated with the passing of a shock wave is different. 

Suppose there is an instantaneous increase in the density at the galactic plane, from $\rho_0$ to $\rho_1$. This would correspond to an instantaneous increase in the vertical oscillation period, from $\omega_0$ to $\omega_1 =  (\rho_1 / \rho_0)^{1/2} \omega_0$.  A star that initially has $z_0$ and $v_0$ will now oscillate in the modified potential. 

As a function of time, its coordinates will be
\begin{eqnarray}
z(t) &=& z_0 \cos (\omega_1 t) + {v_0 \over \omega_1} \sin (\omega_1 t), \\
v(t) &=& -\omega_1 z_0 \sin (\omega_1 t) + {v_0} \cos (\omega_1 t). 
\end{eqnarray} 
The inverse transformation, which we will shortly require is
\begin{eqnarray}
\label{eq:zvToz0v0}
z_0 &=& z (t) \cos (\omega_1 t) - {v_0 \over \omega_1} \sin (\omega_1 t), \\
v_0 &=& \omega_1 z_0 \sin (\omega_1 t) + {v_0} \cos (\omega_1 t). 
\end{eqnarray}

Since the distribution is assumed to be thermalized prior to the instantaneous change in the potential, the probability distribution is Boltzmannian, but relative to the initial potential. Namely, 
\begin{equation}
P(z_0,v_0) = {1\over Z} \exp \left( - {\beta \over 2 } v_0^2 - {\beta \omega_0^2 \over 2} z_0^2 \right). 
\end{equation}
$Z$ is the normalization (and the partition function of the unperturbed distribution) while $\beta$ is the inverse kinematic temperature. If we use the transformation given in eq.~\ref{eq:zvToz0v0}, we find:
\begin{equation}
P(z,v,t) = {1\over Z} \exp \left( - {v_0^2 \over  2 \sigma_{vv}} - {v_0 z_0  \over  \sigma_{vz}} - {z_0^2 \over  2 \sigma_{zz}} \right), 
\end{equation}
with
\begin{eqnarray}
\sigma_{vv}^{-1} &\equiv& \beta \left( \cos^2 \phi +{\omega_0^2 \over \omega_1^2} \sin^2 \phi  \right), \\
\sigma_{vz}^{-1} &\equiv& {\beta \over 2} \left( \omega_1^2 - \omega_0^2 \over  \omega_1^2 \right) \omega_1 \sin (2 \phi), \\
\sigma_{zz}^{-1} &\equiv& \beta \left( \omega_0^2 \cos^2 \phi +\omega_1^2 \sin^2 \phi  \right), 
\end{eqnarray}
and $\phi = \omega_1 t$.

The resulting distribution is a bivariate normal distribution. However, because the covariant term is generally not vanishing (except during multiples of a quarter period), the distribution cannot be described by a thermal distribution. 

To see this, let us calculate the  quadratic moments. We find 
\begin{eqnarray}
\left\langle z^2 \right\rangle &=&  \int z^2 P(z,v,t) dv dz  \\ & = & \nonumber
     {\omega_1 ^2 + \omega_0^2 + \left( \omega_1^2 - \omega_0^2 \right) \cos (2 \phi) \over 2 \beta \omega_0^2 \omega_1^2}, \\
\left\langle v z \right\rangle &=&  \int v z P(z,v,t) dv dz \\ & = & \nonumber - {\left( \omega_1^2 - \omega_0^2 \right) \sin (2 \phi) \over 2 \beta \omega_0^2 \omega_1 }, \\
\left\langle v^2 \right\rangle &=&  \int v^2 P(z,v,t) dv dz \\ & = & \nonumber
{\omega_1^2 + \omega_0^2 - \left( \omega_1^2 - \omega_0^2 \right) \cos (2 \phi) \over 2 \beta \omega_0^2}. 
\end{eqnarray}

A thermal distribution would normally have $\left\langle v z \right\rangle =0$, but not in this case. We can  understand the physical origin of this term by considering the two other terms, $\left\langle z^2 \right\rangle$ and $\left\langle v^2 \right\rangle$. Both oscillate, but in opposite phase of each other. This means that the perturbed distribution oscillates between being puffed up in $z$, with smaller velocities, to having larger velocities but more compact in the $z$ axis. In between the stellar distribution is either getting wider in $z$ or narrower. When it is getting wider, the stars above the plane have a net positive velocity while those below the plane will have a net negative velocity, giving rise to a correlation between $z$ and $v$ and therefore $\left\langle v z \right\rangle >0$. When the stellar distribution is getting more compact, the opposite correlation arises and $\left\langle v z \right\rangle < 0$. This describes ``ringing" motion of the stars. 

Another very interesting implication is that if one uses the thermal distribution to estimate the vertical potential one would be systematically off. For a thermal distribution, the second moments are related to the oscillation frequency $\omega$ through $\left\langle v^2 \right\rangle / \left\langle z^2 \right\rangle = \omega^2$. Thus, if the post-shock distribution would have been thermalized, we would have expected this relation to hold with $\omega_1$. Instead, we find an ``effective" frequency of:
\begin{equation}
\omega_\mathrm{eff}^2 \equiv { \left\langle v^2 \right\rangle \over  \left\langle z^2 \right\rangle } = \left(  \omega_1^2 + \omega_0^2 - \left( \omega_1^2 - \omega_0^2 \right) \cos (2 \phi)  \over \omega_1 ^2 + \omega_0^2 + \left( \omega_1^2 - \omega_0^2 \right) \cos (2 \phi) \right) \omega_1^2. 
\end{equation}
That is, the oscillation frequency that one would infer from a naive comparison between the dispersion in velocity and the dispersion in height would not be the correct one. 

Shortly after the shock passage, while the average density is high, the inferred frequency would then be within the range $\omega_0 \leq \omega_\mathrm{eff} \leq \omega_1^2/\omega_0$, depending on the exact oscillation phase. However, as the background ISM approaches the next spiral arm passage, the background density will decrease, such that inferred frequencies change adiabatically by a factor of $(\omega_0/\omega_1)^2 $, giving a range of inferred frequencies of  $\omega_0^2/\omega_1 \leq \omega_\mathrm{eff} \leq \omega_1$. 

If we have no information on the time since the last spiral shock passage, then a spiral shock jump of $\rho_0 \rightarrow \rho_1$ would therefore imply that any effective density within the following range can be obtained:
\begin{equation}
	\rho_0^2/\rho_1 \leq \rho_\mathrm{eff} \leq \rho_1^2/\rho_0.
	\label{eq:rhorange} 
\end{equation} 

This ambiguity in the determination of the oscillation period (and therefore the density and g potential) is actually deeper. The one dimensional thermal distribution of stars in a harmonic potential is characterized by $\beta$ and $\omega$. These parameters can be derived from measuring two others, such as the dispersion in the velocity and the dispersion in height. If however this distribution is perturbed by a shock, the stellar distribution will be characterized by 4 parameters instead of the above 2. They are $\beta$, characterizing the kinetic temperature of the distribution, $\omega_0$, the oscillation period before the density perturbation, and $\omega_1$, after the perturbation, and last, the phase $\phi = \omega_1 t$ describing how much has the distribution rotated since the perturbation. However, the bivariate Gaussian distribution after the perturbation is characterized by only three variables. Thus, measurement of $\left\langle v^2 \right\rangle$, $\left\langle z^2 \right\rangle$ and $\left\langle vz \right\rangle$ cannot recover the above four parameters without an ambiguity. In particular, a {\em local} measurement of the stellar distribution cannot unambiguously recover the vertical potential if it was perturbed by a spiral arm passage.


\def\asym{{\cal A}}
\def\sgn{{\mathrm{sgn}}}

\section{Asymmetry parameters}
\label{sec:asymetry}

In order to see whether the local distribution of stars includes a signature expected from a spiral arm passage, we can calculate the above quadratic moments of a local set of stars. A useful quantity would be a dimensionless one such that an ${\cal O}(1)$ would imply having a significantly non-thermal distribution, thus, we can define
\begin{eqnarray}
\asym_2 &\equiv& { \left\langle vz \right\rangle \over \sqrt{ \left\langle v^2 \right\rangle \left\langle z^2 \right\rangle }} = { (\omega_1^2 - \omega_0^2) \sin(2 \phi) \over \sqrt{(\omega_1^2 +\omega_0^2)^2 - (\omega_1^2 -\omega_0^2)^2 \cos^2 (2 \phi) }}  \nonumber
\\ &\approx&  {\sin (2 \phi) \over \omega_0} \delta\omega, 
\label{eq:asymA2}
\end{eqnarray}
where $\delta \omega \equiv \omega_1 - \omega_0 \ll \omega_0$. When $\omega_1 = 2 \omega_0$, for example, we find $| \asym_2| \leq 3/5$.

However, there is an advantage in considering lower order moments instead of considering the quadratic ones.  Quadratic moments give a higher weight to faster stars, and to stars located further away from the galactic plane. They are therefore more sensitive to the effect of outliers (the velocity distribution becomes significantly non-Gaussian for stars having $|v| \gtrsim 20~$km/s, corresponding to $z_{max} \sim 200~$pc) and the fact that the Hipparcos F-stars data used below becomes incomplete beyond about 200 pc. These can introduce systematic errors. Moreover, as we see in \S\ref{sec:nonImpulsive}, we expect the asymmetry signal to be larger for smaller amplitude stars (which see a more impulsive change in the galactic potential). Thus, we can define a dimensionless asymmetry parameter that is more sensitive to lower velocity stars, as:
\begin{equation}
\asym_0 \equiv {1 \over Z} \int \sgn(v) \sgn(z) P(z,v,t) dv dz,
\end{equation} 
where $\sgn(x)$ gives $\pm1$ depending on the sign of $x$.
It can be evaluated for the above bivariate distribution to give
\begin{equation}
\asym_0 = - {2 \over \pi} \tan^{-1} \left( (\omega_1^2 - \omega_0^2) \sin (2 \phi) \over 2 \omega_0 \omega_1 \right) \approx - {2 \over \pi} {\sin (2 \phi) \over \omega_0} \delta\omega,
\end{equation} 
where $\delta \omega = \omega_1 - \omega_0 \ll \omega_0$. For the specific case of $\omega_1 = 2 \omega_0$, one has   $ |\asym_0| \leq 0.41$.

\section{A non-impulsive perturbation}
\label{sec:nonImpulsive}

\begin{figure}
\centerline{\includegraphics[width=3in]{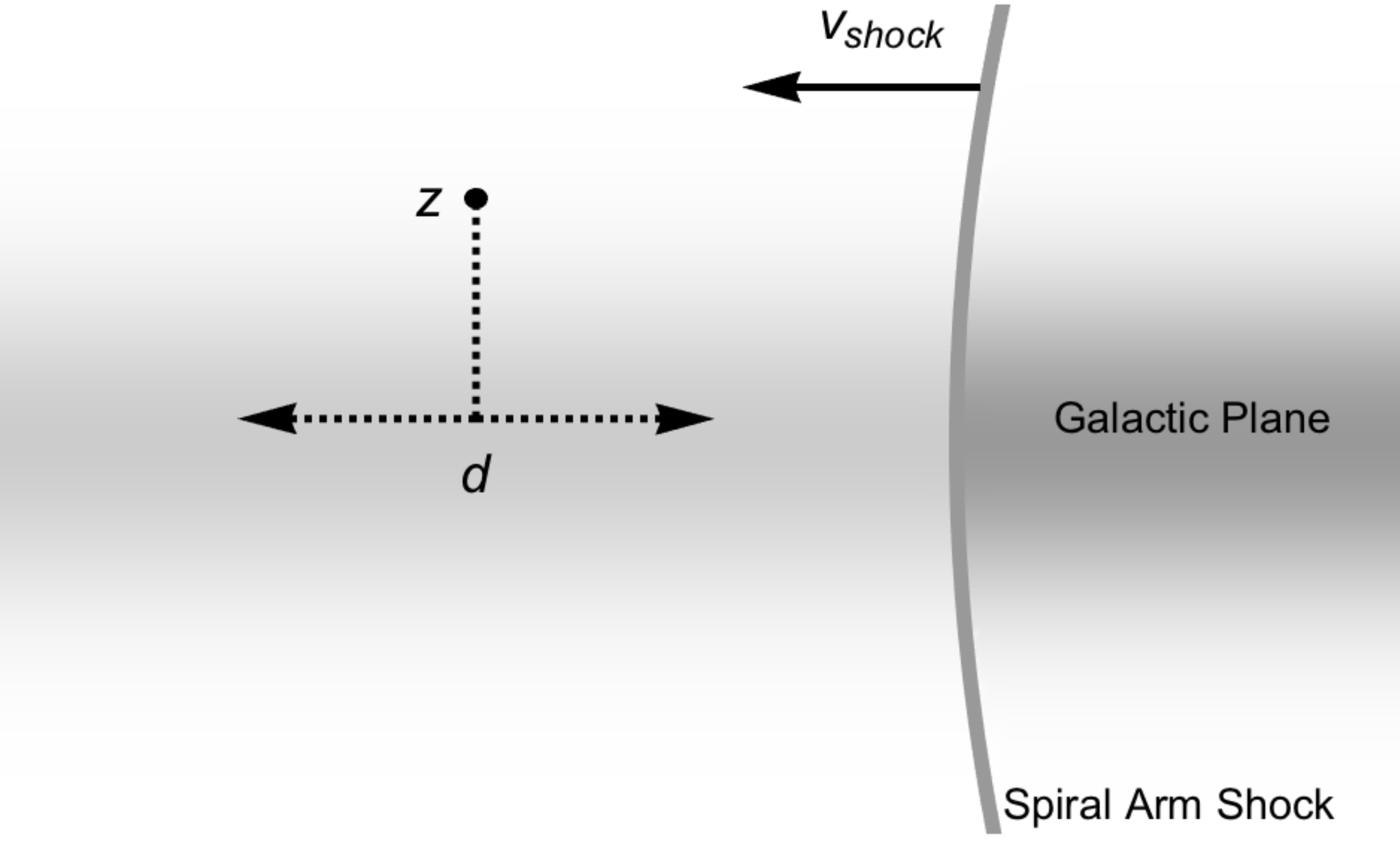}}
\caption{A cartoon of the stellar oscillation with respect to the spiral arm shock. Whether the vertical oscillation can be considered adiabatic or impulsive depends on the ratio between the time it takes the shock, moving with $v_\mathrm{shock}$, to cross the zone $d \sim z_\mathrm{max}$ influencing the oscillation, and the typical oscillation time, given by $1/\omega$.  
}
\label{fig:cartoon}
\end{figure}

We have assumed thus far that the change in the potential was impulsive. However, as can be seen in fig.\ \ref{fig:cartoon}, this requires the potential to change sufficiently fast compared to the oscillation time scale. Since the horizontal extent of the region influencing an oscillation is of order the amplitude of the oscillation $z_\mathrm{max}$, am impulsive potential change would require the shock, moving at $v_\mathrm{shock}$, to cross the region in a time shorter than $\sim 1/\omega$. We therefore define a dimensionless ``impulsiveness" number as
\begin{equation}
{\cal M} \equiv {v_{shock} \over \omega z_{max}},
\end{equation}  
such that ${\cal M} \gg 1$ corresponds to the impulsive limit, while ${\cal M} \ll 1$ corresponds to the adiabatic one.

Since the slowest that the spiral arm can move and still shock the ISM is of order the typical speed of sound, which is $\sim 10$~km/s, we expect that any spiral arm will give rise to at least a modestly impulsive effect, with ${\cal M} \gtrsim 1 $ for modestly young stars having $z_\mathrm{max} \lesssim 100$~pc. More specifically, if we  adopt a 4-armed spiral arm passage frequency of $\sim 145$~Myr \citep{Shaviv:2002}, a pitch angle of $\sim 30^\circ$ \citep{Naoz:2007}, and a vertical frequency $\omega \sim 0.10$~Myr$^{-1}$ \citep{Shaviv:2014}, we find ${\cal M} \sim 4 (z_\mathrm{max} / 100 pc)^{-1}$. 

To see the effect of finite impulsiveness, we need to solve the kinematics of a test star which experiences a jump in the background density. We can do so by writing the force with the gravitational Green's function and employing the symmetry about the disk plane. We therefore have that the force (per unit mass) at an arbitrary coordinate ${\bf x}$ is
\begin{equation}
{\bf F} = \iiint_{|z'|<z_{max}} {\bf f}({\bf x},{\bf x'}) \rho({\bf x}') d{\bf x'}
\end{equation}
where  ${\bf f}(x,x') = G \cdot ({\bf x}-{\bf x'}) / \left| {\bf x}-{\bf x'} \right|^3$ is the Green's function, with $G$ being the gravitational constant. If the shock is located at $x=0$, such that $\rho(x<0) = \rho_0$ and $\rho(x>0) = \rho_1$ then 
\begin{eqnarray}
{\bf F} &=& \rho_0  \iiint_{|z'|<z_{max},x'<0} {\bf f}({\bf x},{\bf x'}) d{\bf x'} \nonumber 
\\ & & + \rho_1  \iiint_{|z'|<z_{max},x'>0} {\bf f}({\bf x},{\bf x'}) d{\bf x'}.
\end{eqnarray}

\begin{figure}
\centerline{\includegraphics[width=\figurewidth]{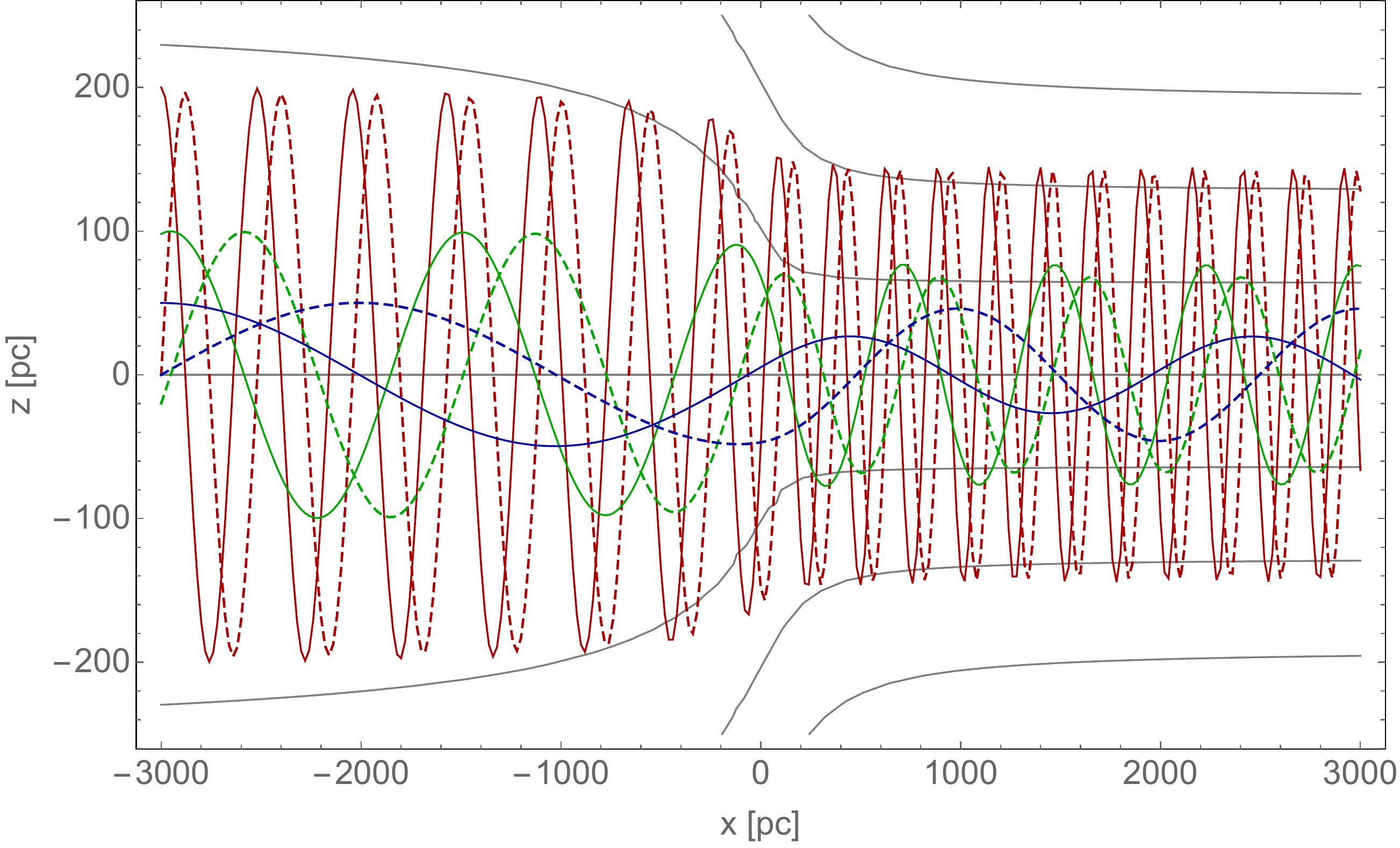}}
\caption{Trajectories of stars starting with a pre-shock density of $\rho_0=0.08$~M$_\odot$/pc$^3$ (left), and a post shock density of $\rho_1 = 4 \rho_0$ (right), for which ${\cal M} = 0.4$ (amplitude of 200~pc in the low density pre-shocked side and an unphysically slow shock velocity of 5 km/s, in red), ${\cal M} = 2.4$ (100~pc, 15 km/s, in green) and ${\cal M} = 12.8$ (50~pc, 40~km/s, in blue). The dashed and solid lines correspond to stars having a different phase in the vertical oscillation. Note that one can see in the high ${\cal M}$ case that the amplitude of two stars having the same initial amplitude is different after the shock. The background contour plot is of equal strength force lines.
}
\label{fig:six-trajectories}
\end{figure}

\begin{figure}
\centerline{\includegraphics[width=\figurewidth]{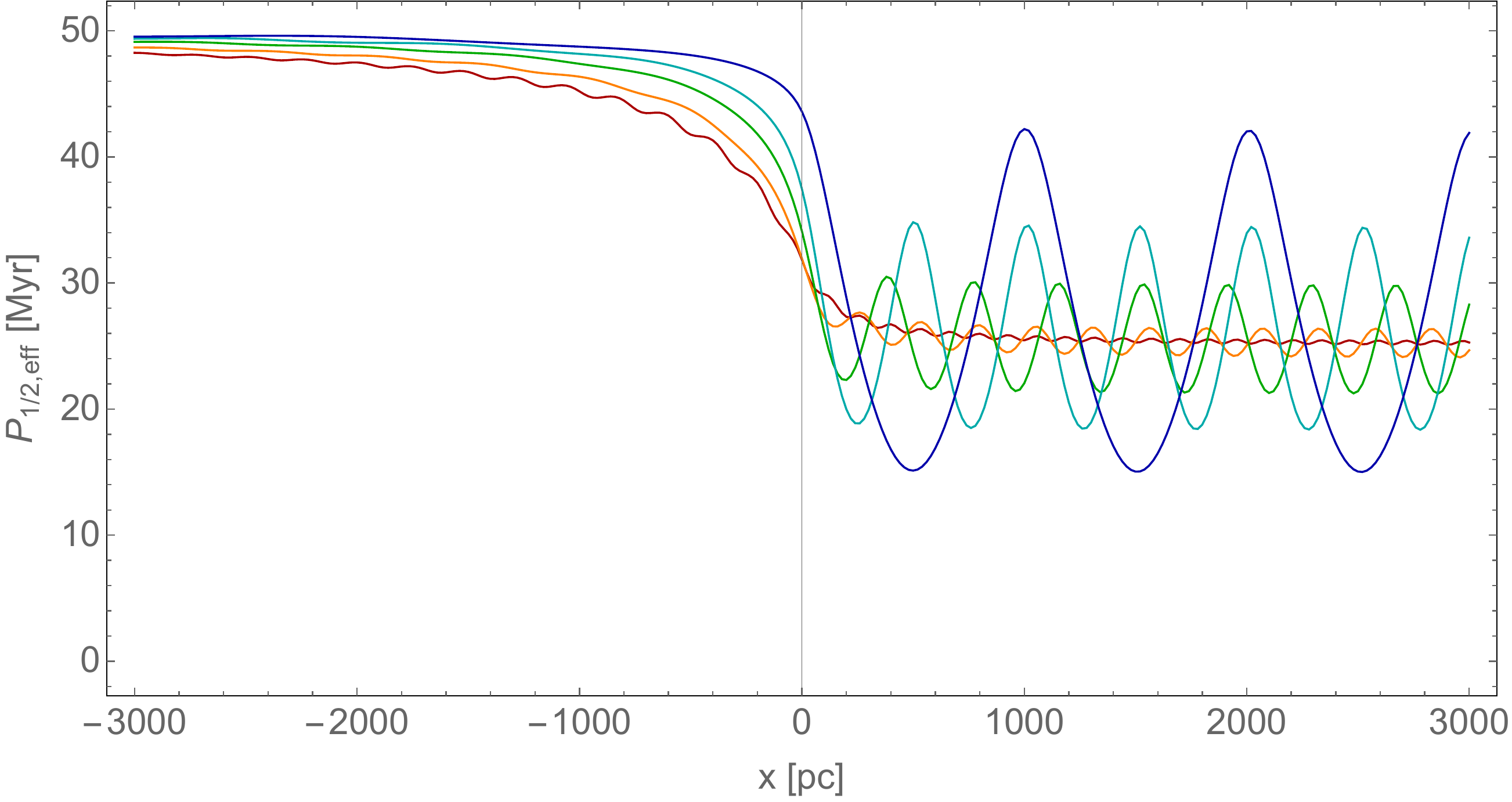}}
\caption{The effective half period of oscillation, $ \pi \left< z^2 \right> / \left< v_z^2 \right>$, inferred from the phase space variance in the populations, for distributions of stars having different ${\cal M}$'s (0.4, 1.05, 2.4, 4.5, 12.8, with 0.4 corresponding to the smallest ``ringing" and 12.8 to the largest).  
}
\label{fig:PhalfRinging}
\end{figure}

\begin{figure}
\centerline{\includegraphics[width=\figurewidth]{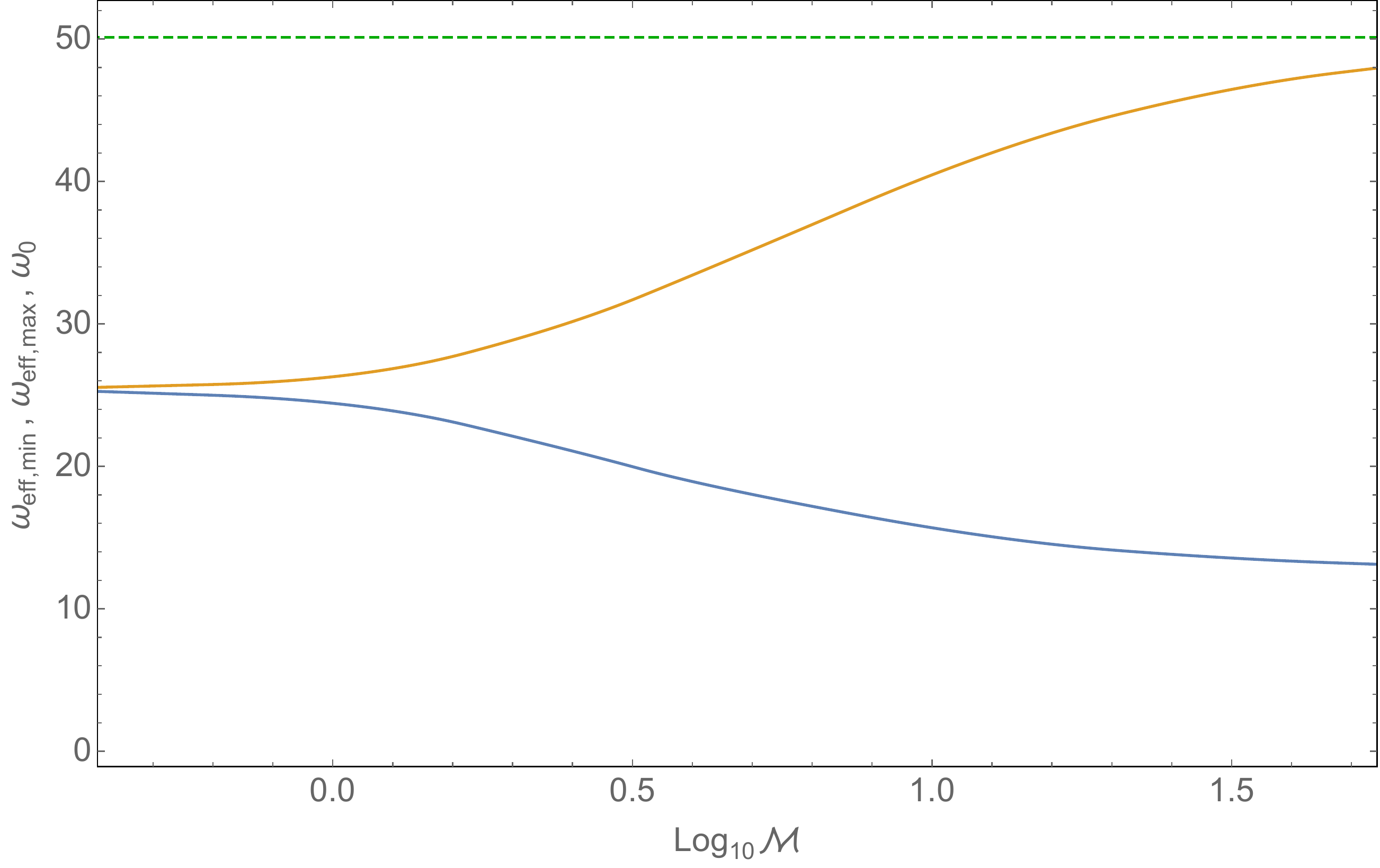}}
\caption{The range of $P_\mathrm{1/2,eff}$ as a function of ${\cal M}$ after the spiral arm passage. The green dashed line represents the maximal value possible for $P_\mathrm{1/2,eff}$ obtained in the impulsive limit. It also corresponds to the pre-shock value of $P_\mathrm{1/2,eff}$. The orange and blue lines correspond to the maximal and minimal values of $P_\mathrm{1/2,eff}$, obtained for stars with different pre-shock phases for a given ${\cal M}$. 
}
\label{fig:omegaeffM}
\end{figure}

\begin{figure}
\centerline{\includegraphics[width=\figurewidth]{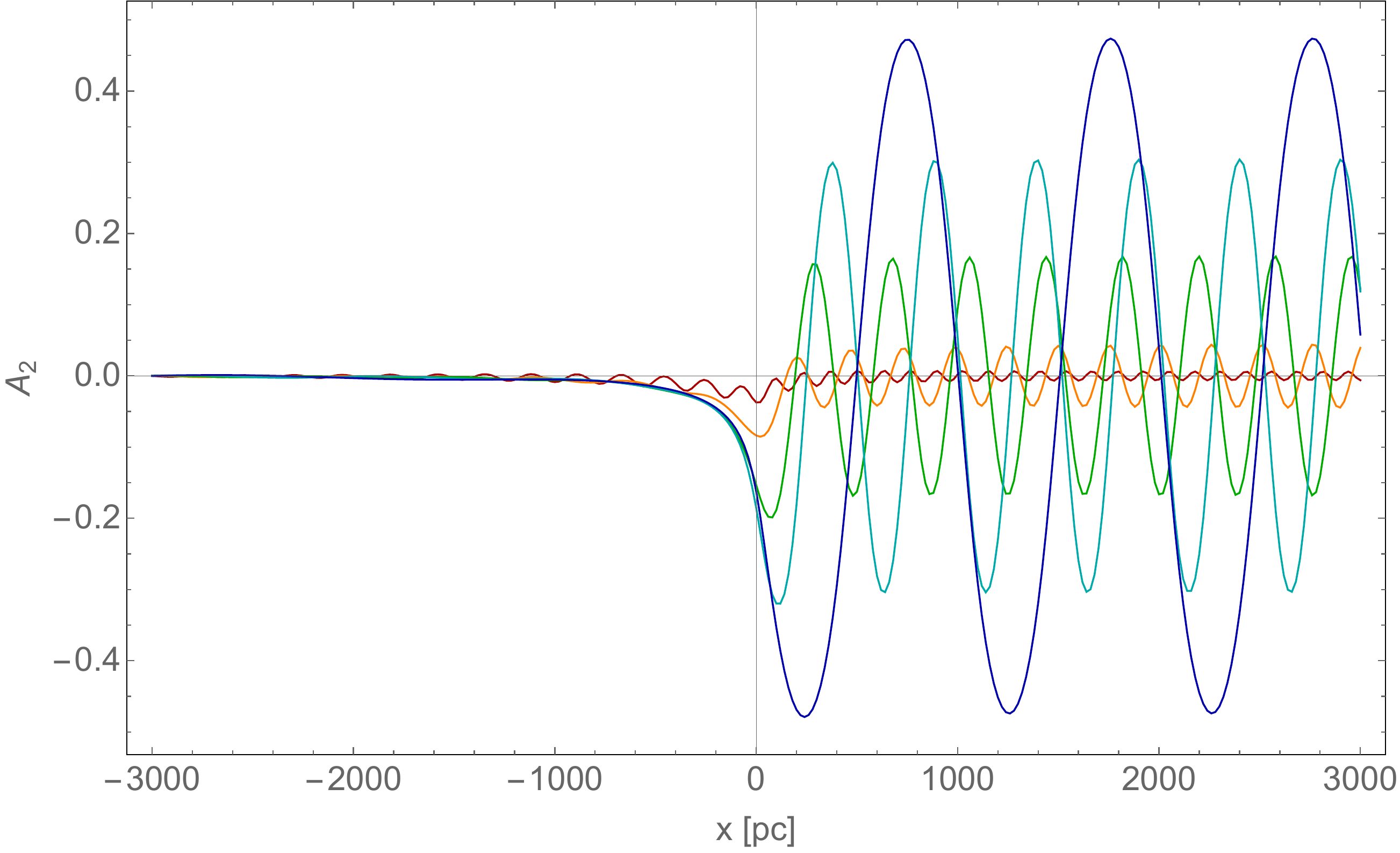}}
\caption{The second moment anisotropy parameter ${\cal A}_2$ for the cases depicted in fig.\ \ref{fig:PhalfRinging}, as a function of the horizontal coordinate. The spiral arm shock is at $x=0$.
}
\label{fig:A2Ringing}
\end{figure}

\begin{figure}
\centerline{\includegraphics[width=\figurewidth]{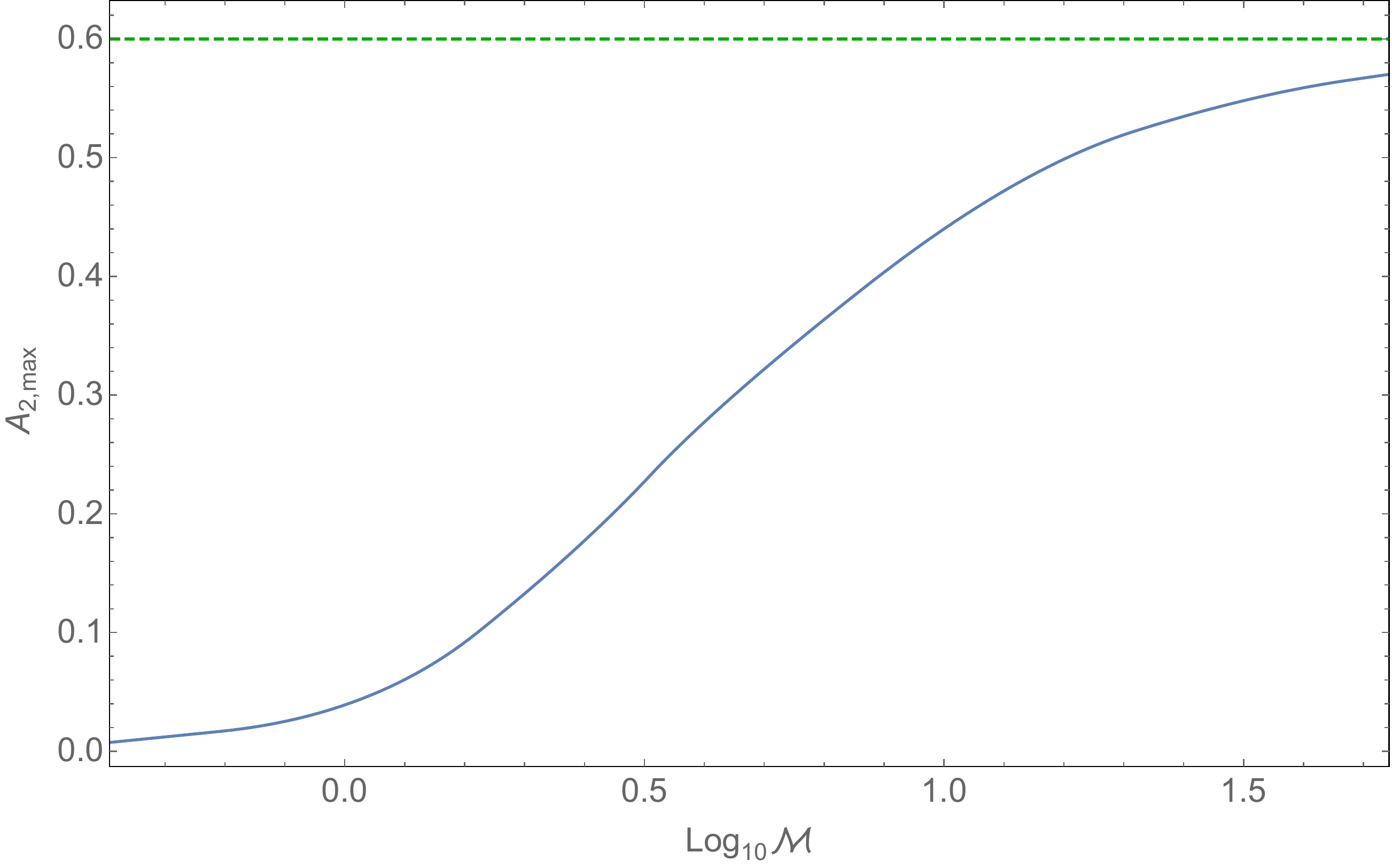}}
\caption{The maximal anisotropy of $\omega_\mathrm{eff}$ as a function of the ${\cal M}$ after the spiral arm passage.
}
\label{fig:maximal-anisotropy}
\end{figure}

Although the general equation, ${d {\bf v} / dt} = {\bf F}$ has a rather complicated solution, we can simplify it considerably by approximating the motion to have a constant velocity in the horizontal direction (such that $v_x =  v_\mathrm{shock} t = \mathrm{const}$), while solving for the vertical direction. Under this approximation, we have ${d v_z / dt} = {{\bf \hat z} \cdot {\bf F}}$ with ${\bf F}$ given above. This equation can then be solved for various cases described in figs.\ \ref{fig:six-trajectories}-\ref{fig:maximal-anisotropy}.

Fig.\ \ref{fig:six-trajectories} describes the solution of ${\cal M}=0.4,2.4$ and $12.8$ cases with $\omega_1 = 2 \omega_0$. Also shown is a contour plot of $F_z(x,z)$. One can see from the figure that the amplitude of the stellar oscillation in the adiabatic limit does not depend on the phase with which the shock catches the star, while in the impulsive limit it is important. This gives rise to the ringing effect in the impulsive limit.

Fig.\ \ref{fig:PhalfRinging} depicts the oscillation half period (i.e., the galactic plane crossing period) that will be inferred by measuring the ratio $\left< z^2 \right> / \left< v_z^2 \right>$. In the adiabatic limit, this inferred half period would correctly change from the real half period in the low density side, to the value in the high density side. However, once we consider impulsive variations in the density, the inferred value down stream from the shock would not necessarily reflect the real vertical period. Instead, it would oscillate around it. For the purely impulsive limit (that is reached only for stars with very small oscillation amplitudes), it could be as high as the upstream period. This result implies that a kinematic measurement of the potential can be systematically wrong. The range of possible half periods for a given ${\cal M}$ is depicted in fig.\ \ref{fig:omegaeffM}.

The ringing effect will manifest itself as an oscillating correlation between $v$ and $z$ of the stars, which can be quantified with the asymmetry parameters. Fig.\ \ref{fig:A2Ringing} demonstrates the oscillation of $\asym_2$ defined in eq.\ \ref{eq:asymA2}. Note that there is a quarter phase difference between the oscillation of $\asym_2$ and the oscillation of $\omega_\mathrm{eff}$ around $\omega_1$. This implies that even if the asymmetry parameter vanishes, we cannot know for certain that ringing is not present and therefore $\omega_\mathrm{eff} \neq \omega_1$. The maximal anisotropy for any ${\cal M}$ is given in fig.\ \ref{fig:maximal-anisotropy}. We see, for example, that one needs ${\cal M} \sim 7$ to get a maximal anisotropy in the oscillations which is two thirds the maximal anisotropy expected in the impulsive limit. This means that finite impulsiveness is important when analyzing the data. It also implies that we can expect a larger anisotropy when considering stars with a smaller amplitude of oscillation.

\section{Real Data: Analysis of A \& F stars}
\label{sec:realdata}

For the data set use the extended Hipparcos-Tycho catalogue and select A and  F stars having kinematic data and for which $-0.2 < B-V<0.6$ and $1.0 < M_V < 2.5$. This includes A0 to F5 stars as a result of the relatively large dispersion in the absolute magnitude / spectral type relation in the Hipparcos dataset \citep{Houk:1997,Jaschek:1998}. Stars in the above spectral range have a typical age ranging from 150 to 1500 Myr. Such stars are on one hand young enough not to have halo contamination, but old enough to thermalize in the vertical potential. They are also sufficiently bright to reach a completeness distance of about 200 pc, which we use as a cutoff for the data. The total number of stars satisfying the above criteria is 5112.

\begin{figure}
\centerline{\includegraphics[width=3.4in]{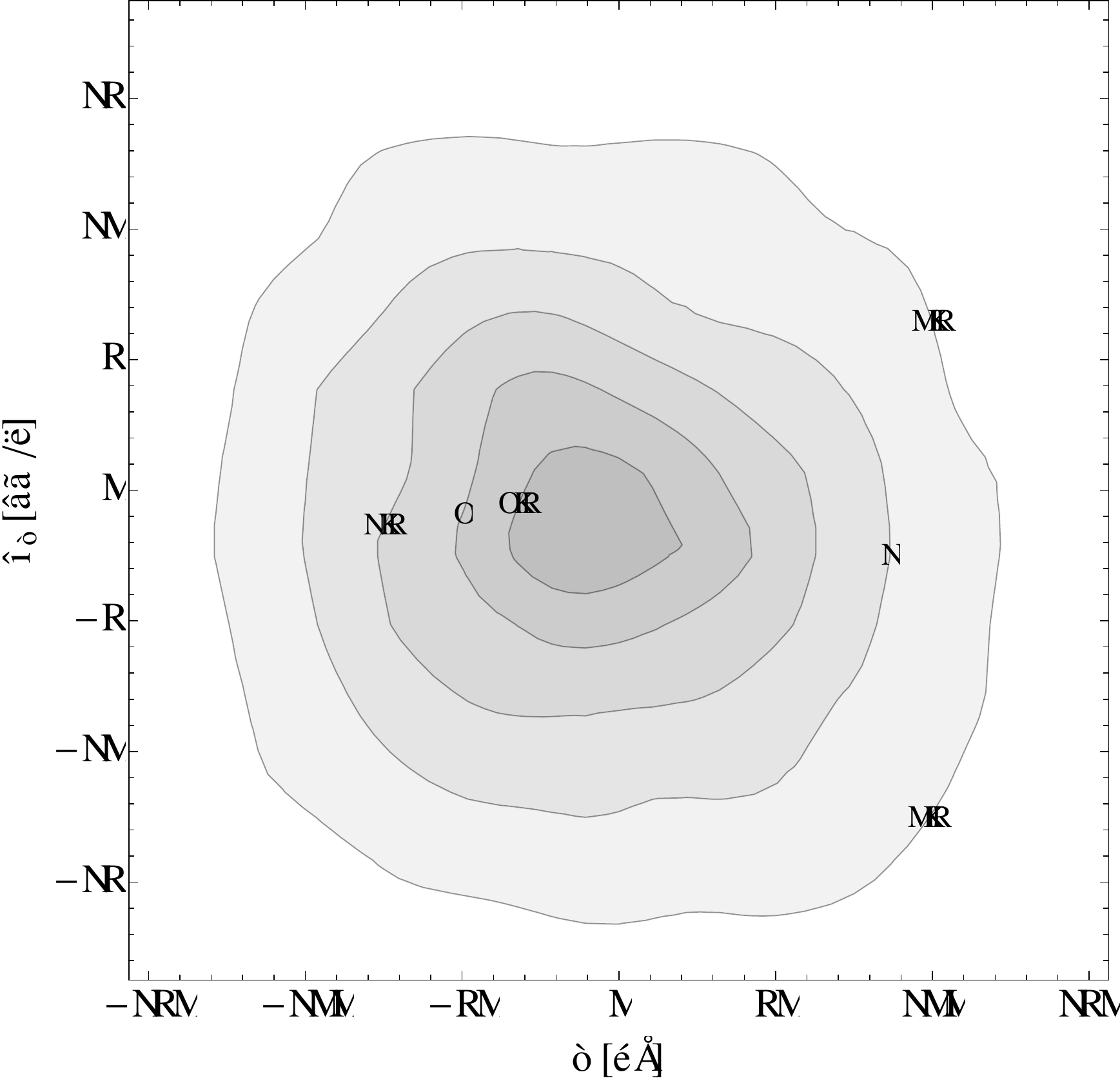}}
\caption{The two dimensional $z-v$ distribution of A and F stars from the Hipparcos catalogue. The contour levels are in units of 1/(pc km/s). There is a lack of a mirror symmetry along the axes ($z \rightarrow -z$ or $v \rightarrow -v$). Instead, there is a rotated symmetry line. As a consequence there is a negative correlation between $z$ and $v$ which describes a contracting distribution.}
\label{fig:2Ddist}
\end{figure}

Fig.\ \ref{fig:2Ddist} depicts the $z - v_z$ distribution of stars. It reveals that the stellar distribution is asymmetrical. Stars having a negative $z$ tend to have a positive velocity while stars having a positive $z$ tend to have a negative vertical velocity. This implies that the distribution is {\em slowly contracting}. The effect is more prominent for stars with a smaller amplitude. This is consistent with the expectations as the smaller amplitude stars experience spiral arm passages more impulsively. 

Using the data, we can calculate the asymmetry parameters, and find 
\begin{eqnarray}
\asym_2 & = & 0.041 \pm 0.016, \\
\asym_0 & = & 0.038 \pm 0.014.
\end{eqnarray}
The error estimate is statistical and based on the variance and the finite sample size. As mentioned above, we expect $\asym_2$ to have a larger systematic error, and a smaller signal. We will therefore use $\asym_0$ below when estimating $\omega_0 / \omega_1$ and from it $\delta\rho/\rho$.

Besides the asymmetry, we can calculate other characteristics of the distribution. We find that $ \sigma_z = {\left\langle z^2 \right \rangle}^{1/2}  =   74.4 \pm 0.7~ \mathrm{pc}$ and $ \sigma_v  {\left\langle v^2 \right \rangle}^{1/2}  =  10.6 \pm 0.2~ \mathrm{km~s^{-1}} $.
These can be used to calculate $\omega_\mathrm{eff}$. However, such a calculation would give an $\omega_\mathrm{eff}$ which could be systematically offsetted. This is because ${\left\langle z^2 \right \rangle}$ could be underestimated due to missing stars at large $z$'s due to completeness of the A and F stars in the Hipparcos data, while the velocity dispersion, especially at large $z$'s can be contaminated by outliers. 

To overcome this problem, we can fit the distribution in $z$ to a gaussian while ignoring data for which $|z|>z_{max}$. This can be repeated for $z_{max}$ between 100 pc and 200 pc. The standard deviation of the truncated Gaussian distribution obtained this way is $\sigma_z = 99\pm 9 ~\mathrm{pc}$. Here the error is the one which encompasses the range of $\sigma_z$ obtained in the fit with different cutoffs. 
We can also calculate the velocity dispersion at the galactic plane, and find that it is $\sigma_v = 8.6 \pm 0.5$~km/s.
 Using these dispersions, we can estimate $\omega_\mathrm{eff}$. It is $ \omega_\mathrm{eff} = {\sigma_v /\sigma_z } =  0.087\pm 0.009~\mathrm{Myr^{-1}} $.
The corresponding effective plane crossing period is then 
$P_{1/2,\mathrm{eff}} \equiv {\pi / \omega_\mathrm{eff}} = 36 \pm 4 ~\mathrm{Myr}$ and the corresponding density at the plane is 
$\rho_0 = 0.135 \pm 0.015$~M$_\odot/$pc$^3$.

Note that even with the fit to the truncated distribution, this value should only be used as a guideline. This is because the distributions are not Gaussians such that a simple analysis is an over simplification. 

A more detailed analysis by \cite{Holmberg:2000} of the kinematics of A \& F stars, which assumes that their ``phases" in phase-space are equilibrated, finds that the density should be $0.103 \pm 0.01$~M$_\odot/$pc$^3$. It was shown however in \cite{Shaviv:2014} that  different cuts of the same data give different estimates for the mass density. Once this systematic uncertainty is considered, a more conservative estimate for the density is $0.12 \pm 0.025$~M$_\odot/$pc$^3$, corresponding to a plane crossing period of $P_{1/2} = 38.3 \pm 3.5$~Myr and an effective frequency of $\omega_\mathrm{eff} = 0.082 \pm 0.008$~Myr$^{-1}$.

The paleoclimatic determination of the plane crossing period is $31.9 \pm 0.6_{sys} \pm 0.6_{stat}$~Myr, such that  $\omega_\mathrm{eff} / \omega_\mathrm{avr} = 0.84\pm0.08$. This average frequency, $\omega_\mathrm{avr}$, should be related to  $\omega_0$ and $\omega_1$ through an appropriate average. In the linear regime for which $\omega_1 - \omega_1  \ll \omega_0$, however, we expect the functional form not to be important, such that we can calculate $\omega_\mathrm{eff} / \omega_\mathrm{avr}$  with for example $\omega_\mathrm{avr} = \sqrt{ \omega_0 \omega_1}$. 

\begin{figure}
\centerline{	
\includegraphics[width=3.2in]{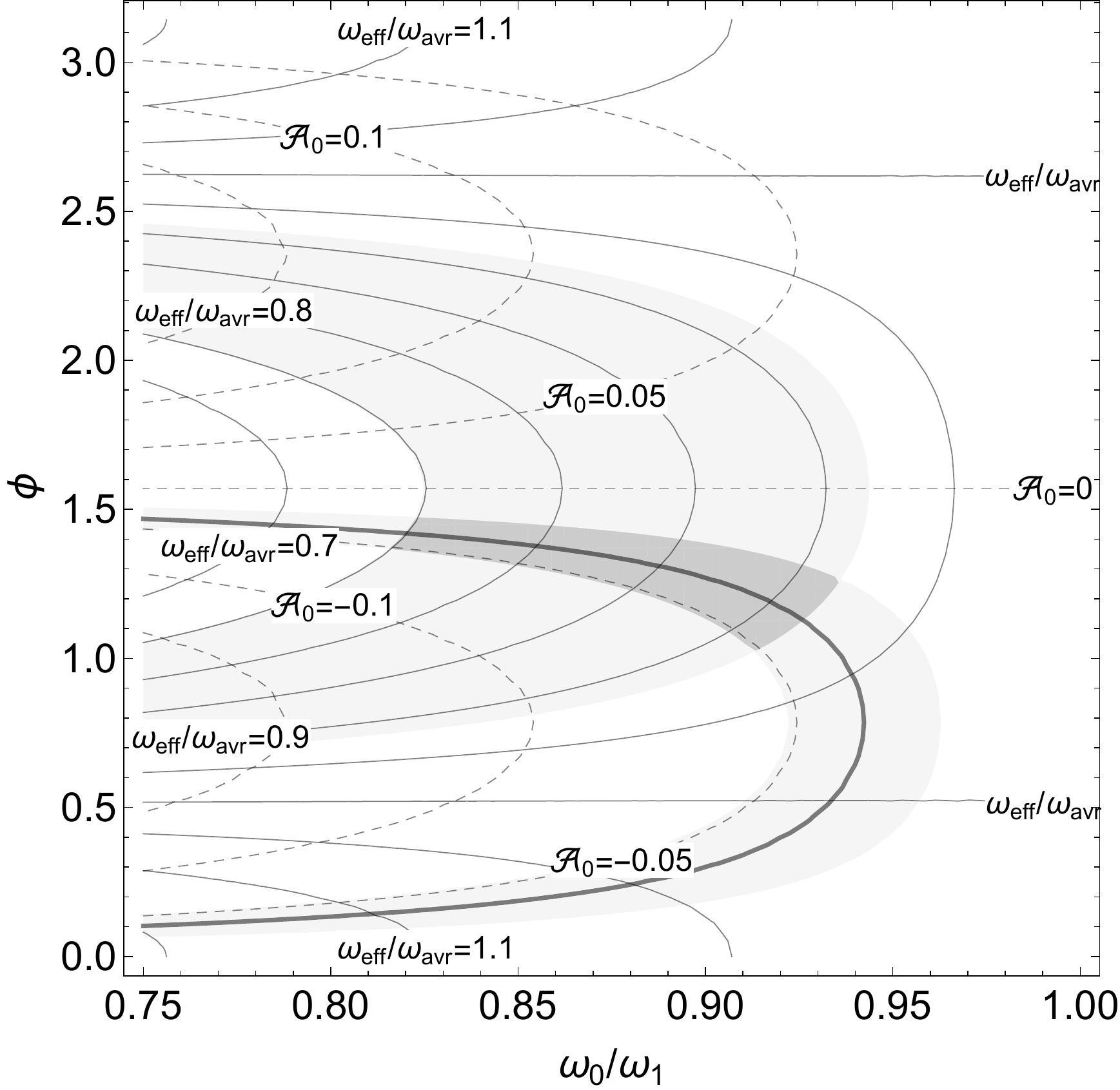}
}
\caption{The asymmetry parameter $\asym_0$ and $\omega_\mathrm{eff}/\omega_{avr}$ (with $\omega_{avr}=\sqrt{\omega_0 \omega_1}$)  as a function of the size of the perturbation, characterized by $\omega_0/\omega_1$, and the oscillation phase $\phi$ since the perturbation has taken place. The lower shaded region corresponds to the measured asymmetry using A and F stars from the Hipparcos catalogue. The upper shaded region corresponds the ratio between the $\omega_\mathrm{eff}$ as derived using the kinematics of A and F stars, and $\omega_\mathrm{eff}$ using paleoclimatic data.  The intersection is highlighted. 
}
\label{fig:OmegaEff}
\end{figure}

Fig.\ \ref{fig:OmegaEff} depicts the asymmetry parameter and the effective frequency $\omega_\mathrm{eff}$ normalized to the frequency average $\sqrt{\omega_0 \omega_1}$.  The shaded regions correspond to either the asymmetry parameter $\asym_0$, as derived from the asymmetry in the velocity/vertical distribution offset of A and F stars, or $\omega_\mathrm{eff}/\omega_\mathrm{avr}$ as inferred from the kinematics of A and F stars and the paleoclimate data.   

One finds from the figure that the consistent frequency jump during the spiral arm passages is $\omega_0 /\omega_1 \approx 0.82-0.93$, corresponding to $\rho_0/\rho_1 \approx 0.67-0.87$. Moreover, the frequency ratios and asymmetry is consistent with a spiral arm shock passage about $(10 \mathrm{~to~} 15 + n \times 31.9)$~Myr, with integer $n$.  There are of course several caveats as elaborated in the discussion. First, finite speed effects would require a larger density jump to give the same asymmetry, requiring a larger density contrast. On the other hand, if the ringing effect dampens slowly, the ringing oscillation grows as the square root of the number of arm passages, thus requiring a smaller density contrast. 

\section{Summary \& Discussion}
\label{sec:discussion}

One of the implicit assumptions built into most kinematic determinations of the mass in the galactic plane is that the stars have random ``phases" while oscillating in their $z-v_z$ phase space. Although it is less restrictive from assuming complete thermalization of the distribution, it was shown in the above analysis that it can lead to systematically wrong determinations of the local dark matter density. This is because abrupt changes to the gravitational potential of the disk arising from the passages of spiral arm shocks imply that the stellar distribution suddenly finds itself with a distribution which is too wide given the velocity dispersion and the new potential. As a consequence, the stellar disk develops a ``ringing" motion as it oscillates around the new equilibrium.

The first interesting aspect of this motion is that it can cause the distribution {\em to appear} to be in phase equilibrium, while in fact it is not. This can give rise to systematically wrong determinations of the galactic mass density.  

However, unless the ringing motion is caught at multiples of quarter of the post-shock oscillation period, the stellar distribution will exhibit a signature in the form of a phase space asymmetry. Stars moving upwards will have a different average $z$ from stars moving downwards, or alternatively, stars above the plane will have an average velocity that is different from stars below the plane. 

Such a signature exists in the nearby A and F-stars found in the extended Hipparcos catalogue at the $2.5\sigma$ confidence level. Specifically, the present stellar distribution within $z \lesssim 100$~pc is contracting towards the galactic plane and the dimensionless asymmetry parameter is ${\cal A} \sim 0.05$, which implies that the kinematically determined oscillation frequency is unknown to at least ${\cal O}(0.1)$, and the inferred density unknown at the ${\cal O}(0.2)$ level. However, because the local kinematic data is insufficient to determine the shock parameters (strength and passage time), without additional information to remove the degeneracy, local kinematic determinations of the mass density have an intrinsic uncertainty. 

The effect can also explain the discrepancy pointed out by \cite{Shaviv:2014}, in which the inferred oscillation frequency of stars with a {\em smaller} vertical amplitude is smaller than the inferred oscillation frequency of stars reaching larger heights, as if the average density seen by stars with a larger amplitude is larger. Such apparently unphysical behavior is explained by the effect since it dominates the smaller amplitude stars as they experience a more abrupt effect by the shock wave.

In addition to the study of stars further away in the vertical direction, the parameter degeneracy can also be alleviated by studying stars further away in the galactic disk. The upcoming GAIA astrometric catalogue \citep{GAIA} will not only be able to measure the local spiral arm signature much better, it will be able to measure the asymmetry signature as a function of horizontal distance. This will not only allow the determination of the mass density without degeneracy, it will also provide the direction of the spiral arm, its amplitude and the time since the last passage (i.e., $\phi$).

Until then, it is possible to extract more information if we combine the kinematic data with the paleoclimatic determination of the long term oscillation period \citep{Shaviv:2014}. By comparing the latter to the ``effective" oscillation frequency from stellar kinematics \citep{Holmberg:2000,korchagin:2003,Shaviv:2014}, one finds that the spiral arm density jump should correspond to about $\delta \rho /\rho \sim 0.15 $, and that the recent spiral arm passage took place about $(10 \mathrm{~to~} 15 + n \times 31.9)$~Myr ago. However there are several caveats.

First, this analysis assumes the sudden approximation. This is certainly true for the stars closest to the galactic plane, but it starts to be a less-than-ideal approximation for stars further away. For the median star, with $z_\mathrm{max} \sim 75$~pc, we expect about 2/3's of the asymmetry predicted under the sudden approximation limit. 

On the other hand, the analysis also assumes that any ringing is the result of a single spiral arm passage. If however the decay time of the ringing is longer than the typical time between passages, the amplitude will be larger, and the phase not a precise indication of the last arm passage. It is therefore interesting to consider the different mechanisms that may govern the decay.

First, nonlinearity of the harmonic potential will imply that stars with different amplitudes will have a somewhat different oscillation frequency $\omega(z)$, which will phase mix the stars. According to \cite{Bahcall:1985}, we can expect this to be a $\delta \omega_{NL} / \omega  \sim 5$\% effect on stars within $\sim 100$~pc. To smear out the ringing effect, a typical mixing of half the plane crossing period should be accumulated, i.e., the nonlinearity mixing time is roughly
$ \tau_{NL} \sim \left( P_{1/2} / 2\right)  \left( \delta \omega_{NL} / \omega \right)^{-1} \sim 300 ~\mathrm{Myr}$. The damping time scale will be shorter for stars with larger amplitudes and longer for smaller amplitudes.  

Diffusion in the phase space  describing the vertical motion will also phase mix the stars. This diffusion is responsible for slowly increasing the velocity dispersion, and it can be described with a diffusion coefficient that is about $D_w \approx 2.5 \times 10^{-7}$km~s$^{-1}$~yr$^{-1}$ \citep{Wielen:1977}. The diffusion in velocity will give rise to a dispersion in the oscillation phase, which is of order $\sigma_\phi \sim \sqrt{D_w t/2} / \sigma_W$. To disperse the stars by half the plane crossing period, i.e., to get $\sigma_\phi \sim \pi /2$, one has to wait typically $\tau_{D} \sim 10^{9}$yr for $\sigma_W \sim 7$~km~s$^{-1}$. Here the phase mixing is faster for smaller amplitude stars, but given the typically longer time scales, the effect is probably not dominant. 

Another relevant effect that should damp the ringing is phase mixing due to the radial dispersion of stars---the local population of stars is a mix coming from different galactic radii each having a somewhat different vertical oscillation period.  If  the horizontal velocity dispersion is $\sigma_{U}\sim 10~$km/s, then the gyration center of these stars will have a dispersion of order $\sigma_{R} \sim \sigma_{U}/\kappa$. Here $\kappa$ is the radial epicyclic frequency which can be derived from Oort's constants. Since there is a radial gradient in the density $\rho \propto \exp(r/L)$, stars having different gyration centers will have a different average vertical oscillation period. The variation in $\omega$ of the vertical oscillation is therefore roughly
\begin{equation}
{\delta \omega \over \omega} \sim {1 \over 2}{\delta \rho \over \rho} \sim {\sigma_{U} \over 2 \omega_\kappa L} = {\sigma_{U} P_\kappa \over 4 \pi L} \sim 0.03,
\end{equation}
where we have taken $L \sim 4~$kpc, and $P_\kappa \sim 150~$Myr. To washout the kinematic signature, the phases have to be mixed by about half the plane crossing period. Thus, the damping from the radial velocity dispersion is 
 \begin{equation}
\tau_U \sim {P_{1/2} \over 2} \left( \delta \omega \over \omega \right)^{-1} \sim 500 ~\mathrm{Myr}.
\end{equation} 
Although this process does not depend directly on the vertical amplitude of the stars, older populations having a larger amplitude will also have a larger radial dispersion, giving rise to faster mixing.

Given the above processes, it is clear that a typical time scale of a few 100 Myr should characterize the phase mixing and damping of the vertical ringing. Since spiral arm passages occur over this time scale (e.g., every 150 Myr, see \citealt{Shaviv:2002}), we should expect to accumulate around $n \sim 2$ to $3$ spiral arm passages. The  arm passages should occur at random phases within the vertical oscillation, the typical amplitude should be of order $\sqrt{n} \sim 1.5$ times larger than the amplitude expected from one spiral arm passage. 

Clearly then, the exact kinematic asymmetry depends on comparable but competing effects. The non-instantaneous spiral arm passage tends to decrease the asymmetry  but finite number of arm passages will increase it. Thus, we can conclude that a spiral arm signature is present in the kinematic data, but cannot exactly ascertain the spiral arm parameters, at least, not until the GAIA catalogue will be published. It also implies that trying to measure the exact density at the galactic plane can lead to systematically wrong values.   

\section*{Acknowledgements}
Many thanks Matias Zaldarriaga and Scott Tremaine for fruitful discussions. This research project was also supported by the
I-CORE Program of the Planning and Budgeting Committee and the Israel Science Foundation (center 1829/12) and by ISF grant no.\ 1423/15, as well as the IBM Einstein Fellowship at the Institute for Advanced Study.

\def\aj{Astron.\ J.}
\def\na{N.\ Astron.}
\def\apj{Ap.\ J.}
\def\apjl{Ap.\ J.\ Lett.}
\def\mnras{Mon.\ Not.\ Roy.\ Astro.\ Soc.}
\def\aap{Astron.\ Astrophys.}
\def\araa{Ann.\ Rev.\ Astron.\ Astrophys.}
\def\pasj{Pub.\ Astron.\ Soc.\ Japan}
\def\pre{Phys.\ Rev.\ E}
\def\apss{Astrophys.\ Sp.\ Sc.}
\def\solphys{Solar Phys.}
\def\nat{Nature}

\bibliography{SpiralSignatureRefs}

\begin{thebibliography}{}
\expandafter\ifx\csname natexlab\endcsname\relax\def\natexlab#1{#1}\fi

\bibitem[{{Antoja} {et~al.}(2011){Antoja}, {Figueras}, {Romero-G{\'o}mez},
  {Pichardo}, {Valenzuela}, \& {Moreno}}]{Antoja:2011}
{Antoja}, T., {Figueras}, F., {Romero-G{\'o}mez}, M., {et~al.} 2011, \mnras,
  418, 1423

\bibitem[{{Bahcall} \& {Bahcall}(1985)}]{Bahcall:1985}
{Bahcall}, J.~N., \& {Bahcall}, S. 1985, \nat, 316, 706

\bibitem[{{Binney} \& {Tremaine}(1987)}]{Binney:1987}
{Binney}, J., \& {Tremaine}, S. 1987, Galactic dynamics (Princeton University
  Press)

\bibitem[{{Blitz} {et~al.}(1983){Blitz}, {Fich}, \& {Kulkarni}}]{Blitz:1983}
{Blitz}, L., {Fich}, M., \& {Kulkarni}, S. 1983, Science, 220, 1233

\bibitem[{{Bobylev} \& {Bajkova}(2015)}]{Bobylev:2015}
{Bobylev}, V.~V., \& {Bajkova}, A.~T. 2015, \mnras, 447, L50

\bibitem[{{Dame} {et~al.}(2001){Dame}, {Hartmann}, \& {Thaddeus}}]{Dame:2001}
{Dame}, T.~M., {Hartmann}, D., \& {Thaddeus}, P. 2001, \apj, 547, 792

\bibitem[{{Debattista}(2014)}]{Debattista:2014}
{Debattista}, V.~P. 2014, \mnras, 443, L1

\bibitem[{{Fan} {et~al.}(2013){Fan}, {Katz}, {Randall}, \& {Reece}}]{Fan:2013}
{Fan}, J., {Katz}, A., {Randall}, L., \& {Reece}, M. 2013, Physical Review
  Letters, 110, 211302

\bibitem[{{Faure} {et~al.}(2014){Faure}, {Siebert}, \& {Famaey}}]{Faure:2014}
{Faure}, C., {Siebert}, A., \& {Famaey}, B. 2014, \mnras, 440, 2564

\bibitem[{Garbari {et~al.}(2011)Garbari, Read, \& Lake}]{Garbari:2011}
Garbari, S., Read, J.~I., \& Lake, G. 2011, Mon.\ Not.\ Roy.\ Astro.\ Soc.,
  416, 2318

\bibitem[{{Garrido Pesta{\~n}a} \& {Eckhardt}(2010)}]{Pestana:2010}
{Garrido Pesta{\~n}a}, J.~L., \& {Eckhardt}, D.~H. 2010, \apjl, 722, L70

\bibitem[{{Georgelin} \& {Georgelin}(1976)}]{Georgelin:1976}
{Georgelin}, Y.~M., \& {Georgelin}, Y.~P. 1976, \aap, 49, 57

\bibitem[{Holmberg \& Flynn(2000)}]{Holmberg:2000}
Holmberg, J., \& Flynn, C. 2000, Mon.\ Not.\ Roy.\ Astro.\ Soc., 313, 209

\bibitem[{{Houk} {et~al.}(1997){Houk}, {Swift}, {Murray}, {Penston}, \&
  {Binney}}]{Houk:1997}
{Houk}, N., {Swift}, C.~M., {Murray}, C.~A., {Penston}, M.~J., \& {Binney},
  J.~J. 1997, in ESA Special Publication, Vol. 402, Hipparcos - Venice '97, ed.
  R.~M. {Bonnet}, E.~{H{\o}g}, P.~L. {Bernacca}, L.~{Emiliani}, A.~{Blaauw},
  C.~{Turon}, J.~{Kovalevsky}, L.~{Lindegren}, H.~{Hassan}, M.~{Bouffard},
  B.~{Strim}, D.~{Heger}, M.~A.~C. {Perryman}, \& L.~{Woltjer}, 279--282

\bibitem[{{Jaschek} \& {Gomez}(1998)}]{Jaschek:1998}
{Jaschek}, C., \& {Gomez}, A.~E. 1998, \aap, 330, 619

\bibitem[{{Korchagin} {et~al.}(2003){Korchagin}, {Girard}, {Borkova},
  {Dinescu}, \& {van Altena}}]{korchagin:2003}
{Korchagin}, V.~I., {Girard}, T.~M., {Borkova}, T.~V., {Dinescu}, D.~I., \&
  {van Altena}, W.~F. 2003, \aj, 126, 2896

\bibitem[{Lin \& Bertin(1995)}]{Lin:1995}
Lin, C.~C., \& Bertin, G. 1995, Annals of the New York Academy of Sciences,
  773, 125

\bibitem[{{Lin} \& {Shu}(1964)}]{Lin:1964}
{Lin}, C.~C., \& {Shu}, F.~H. 1964, \apj, 140, 646

\bibitem[{{McKee} {et~al.}(2015){McKee}, {Parravano}, \&
  {Hollenbach}}]{McKee:2015}
{McKee}, C.~F., {Parravano}, A., \& {Hollenbach}, D.~J. 2015, \apj, 814, 13

\bibitem[{{Mongui{\'o}} {et~al.}(2015){Mongui{\'o}}, {Grosb{\o}l}, \&
  {Figueras}}]{Monguio:2015}
{Mongui{\'o}}, M., {Grosb{\o}l}, P., \& {Figueras}, F. 2015, \aap, 577, A142

\bibitem[{{Naoz} \& {Shaviv}(2007)}]{Naoz:2007}
{Naoz}, S., \& {Shaviv}, N.~J. 2007, \na, 12, 410

\bibitem[{Oort(1960)}]{Oort:1960}
Oort, J.~H. 1960, Bull. Astron. Inst. Netherlands, 15, 45

\bibitem[{{Perryman} {et~al.}(2001){Perryman}, {de Boer}, {Gilmore}, {H{\o}g},
  {Lattanzi}, {Lindegren}, {Luri}, {Mignard}, {Pace}, \& {de Zeeuw}}]{GAIA}
{Perryman}, M.~A.~C., {de Boer}, K.~S., {Gilmore}, G., {et~al.} 2001, \aap,
  369, 339

\bibitem[{{Pomp{\'e}ia} {et~al.}(2011){Pomp{\'e}ia}, {Masseron}, {Famaey}, {van
  Eck}, {Jorissen}, {Minchev}, {Siebert}, {Sneden}, {L{\'e}pine}, {Siopis},
  {Gentile}, {Dermine}, {Pasquato}, {van Winckel}, {Waelkens}, {Raskin},
  {Prins}, {Pessemier}, {Hensberge}, {Fr{\'e}mat}, {Dumortier}, \&
  {Bienaym{\'e}}}]{Pompia:2011}
{Pomp{\'e}ia}, L., {Masseron}, T., {Famaey}, B., {et~al.} 2011, \mnras, 415,
  1138

\bibitem[{{Quillen} \& {Minchev}(2005)}]{Quillen:2005}
{Quillen}, A.~C., \& {Minchev}, I. 2005, \aj, 130, 576

\bibitem[{{Randall} \& {Reece}(2014)}]{Randall:2014}
{Randall}, L., \& {Reece}, M. 2014, Physical Review Letters, 112, 161301

\bibitem[{{Roberts}(1972)}]{Roberts:1972}
{Roberts}, Jr., W.~W. 1972, \apj, 173, 259

\bibitem[{S{\'{a}}nchez-Salcedo {et~al.}(2011)S{\'{a}}nchez-Salcedo, Flynn, \&
  Hidalgo-G{\'{a}}mez}]{Salcedo:2011}
S{\'{a}}nchez-Salcedo, F.~J., Flynn, C., \& Hidalgo-G{\'{a}}mez, A.~M. 2011,
  {ApJ}, 731, L35

\bibitem[{{Shaviv}(2003)}]{Shaviv:2002}
{Shaviv}, N.~J. 2003, \na, 8, 39

\bibitem[{{Shaviv} {et~al.}(2014){Shaviv}, {Prokoph}, \&
  {Veizer}}]{Shaviv:2014}
{Shaviv}, N.~J., {Prokoph}, A., \& {Veizer}, J. 2014, Scientific Reports, 4,
  6150

\bibitem[{Siebert {et~al.}(2012)Siebert, Famaey, Binney, Burnett, Faure,
  Minchev, Williams, Bienaym{\'{e}}, Bland-Hawthorn, Boeche, Gibson, Grebel,
  Helmi, Just, Munari, Navarro, Parker, Reid, Seabroke, Siviero, Steinmetz, \&
  Zwitter}]{Siebert:2012}
Siebert, A., Famaey, B., Binney, J., {et~al.} 2012, Monthly Notices of the
  Royal Astronomical Society, 425, 2335

\bibitem[{{Stothers}(1998)}]{Stothers:1998}
{Stothers}, R.~B. 1998, \mnras, 300, 1098

\bibitem[{{Vallee}(1995)}]{Vallee:1995}
{Vallee}, J.~P. 1995, \apj, 454, 119

\bibitem[{{Wielen}(1977)}]{Wielen:1977}
{Wielen}, R. 1977, \aap, 60, 263

\end{thebibliography}

\end{document}